\documentclass{appolb}
\usepackage{graphicx}

\usepackage[
	backend=bibtex8,
	bibencoding=ascii,
	language=auto,
	style=numeric-comp,
	sorting=none,
	maxbibnames=10,
	backref=false,
	natbib=true 
]{biblatex}
\usepackage{color}

\begin{document}
\title{Bayesian quantification of the Quark-Gluon Plasma: 
Improved design and closure demonstration%
\thanks{Presented at the XXIXth International Conference on Ultra-relativistic Nucleus-Nucleus Collisions (Quark Matter 2022). }
}
\author{Matthew Heffernan$^\mathrm{a}$\thanks{Speaker}, Charles Gale$^\mathrm{a}$, Sangyong Jeon$^\mathrm{a}$, Jean-Fran\c{c}ois Paquet$^\mathrm{b}$\\
\address{$^\mathrm{a}$Department of Physics, McGill University, Montr\'eal, QC, H3A 2T8, Canada\\$^\mathrm{b}$Department of Physics, Duke University, Durham, NC 27708, USA}}

\maketitle
\begin{abstract}
We present a demonstration of the design sampling and closure for the first comprehensive Bayesian model-to-data comparison of heavy-ion measurements with IP-Glasma initial conditions, in which we combine with state-of-the-art hydrodynamics (MUSIC), particlization (iS3D), and transport (SMASH).
We further introduce a systematically-improvable method of sampling design points with better projection properties to explore the parameter space of the model for the first time in heavy-ion collisions. 
\end{abstract}

\section{Introduction}
The strongly-interacting matter produced in ultra-relativistic heavy ion collisions is successfully modeled with a pre-equilibrium phase, followed by viscous hydrodynamics, particlization, and a hadronic afterburner with the totality of the collision taking place in $\mathcal{O}(10)$ fm/c
\cite{Schenke:2021mxx}. 
Bayesian analysis techniques are increasingly applied to constrain the physical properties of this matter and to quantify their uncertainty (c.f. \cite{SIMSPRL,SIMSPRC} and references therein). 
These provide a systematic, methodical way to solve the statistical inverse problem. 
Due to computational cost, surrogate models are trained on event-by-event hybrid simulations at specific choices of input parameters, called design points, chosen with space-filling sampling strategies. 

IP-Glasma is a pre-equilibrium model in the Color Glass Condensate framework where the gluon field evolution obeys the classical Yang-Mills equation \cite{IP-Glasma}. 
In the hydrodynamic phase, (2+1)D MUSIC \cite{Schenke:2010nt} implements second-order viscous hydrodynamics until the medium  cools to the particlization temperature, where particles become the pertinent degrees of freedom. 
The transition from hydrodynamics to particles is implemented using the 14-moment viscous corrections in iS3D \cite{iS3D}. 
Finally, the particles are evolved from the freezeout surface using SMASH \cite{smash}. The viscous parameters are  parameterized as in Ref. \cite{SIMSPRC} with similar Bayesian priors.

In what follows, we discuss a different, improved approach to space-filling designs for heavy ion collisions before demonstrating self-consistency (closure) tests with varying fractions of the design space. 
The aim of this analysis is to both illustrate the self-consistency of a hybrid model with an IP-Glasma initial stage and to introduce an improved approach to design space sampling that allows for reliable intermediary analysis. These first steps establish the foundation of an upcoming complete study. 
\vspace{-0.1in}
\section{Computer Experiment Design}
\begin{figure}[htb!]
	\centering
	\includegraphics[width=12.5cm]{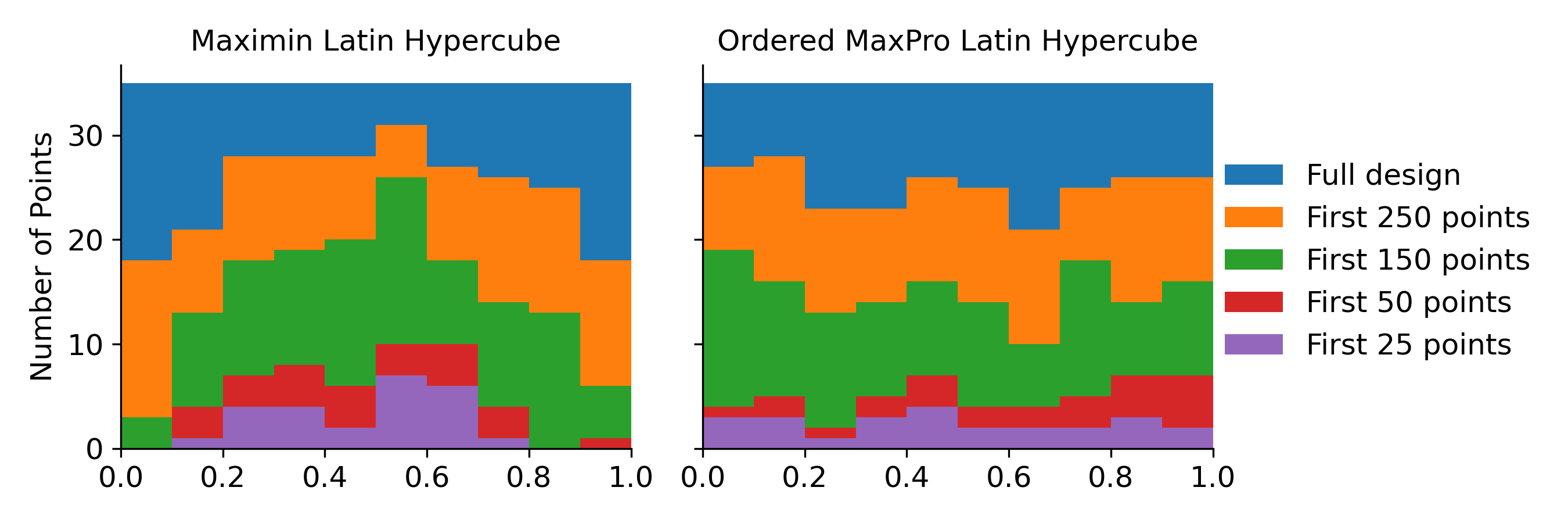}
	\caption{Subsequent subsets from a one-dimensional projection of the unit 11-dimensional Maximum Projection design with 350 design points compared to similar subsets for a maximin Latin hypercube algorithm.}
	\label{fig:maxpro-vs-maximin-1d}
\end{figure}

Previous Bayesian works in heavy ion collisions have used a Latin hypercube \cite{https://doi.org/10.1002/sam.11414} that maximizes the minimum Euclidean distance between points in $d$-dimensional space 
However, some parameters in a study may not be well-constrained and the remainder form an active subspace. 
To ensure that the coverage of the active subspace is also considered, a design that maximizes all projections of the $d$-dimensional design space is desirable. 
An implementation of this is a Maximum Projection design \cite{MaxPro}. These designs minimize the Maximum Projection criterion for points $x$,
\begin{equation}
	\min _{D} \psi(D)=\left\{\frac{1}{\left(\begin{array}{c}
		n \\\vspace{-0.1in}
		2
		\end{array}\right)} \sum_{i=1}^{n-1} \sum_{j=i+1}^{n} \frac{1}{\prod_{l=1}^{p}\left(x_{i l}-x_{j l}\right)^{2}}\right\}^{1 / p}.
\end{equation}
$p$ is the number of dimensions of the space and $n$ are the number of points in the sample space. 
$\psi(D)=\infty$ for $x_{il}=x_{jl}$, $i\neq j$. 
These points may also be ordered such that the first $N$ points are the subset of the full design which minimizes $\psi(D)$ given the constraint that only $N$ points may be chosen. 
This ensures that partially-completed designs can be used to reliably train surrogate models as the design more rapidly achieves coverage of the design space. 
This is shown in Fig.~\ref{fig:maxpro-vs-maximin-1d}, which shows that the currently popular maximin algorithm
prioritizes centrally-clustered points until the design is complete while the ordered Maximum Projection Latin hypercube features more even partial subsets.
\vspace{-0.1in}
\section{Self-consistency testing}
\begin{figure}[!htb]
    \centering
    \includegraphics[width=12.5cm]{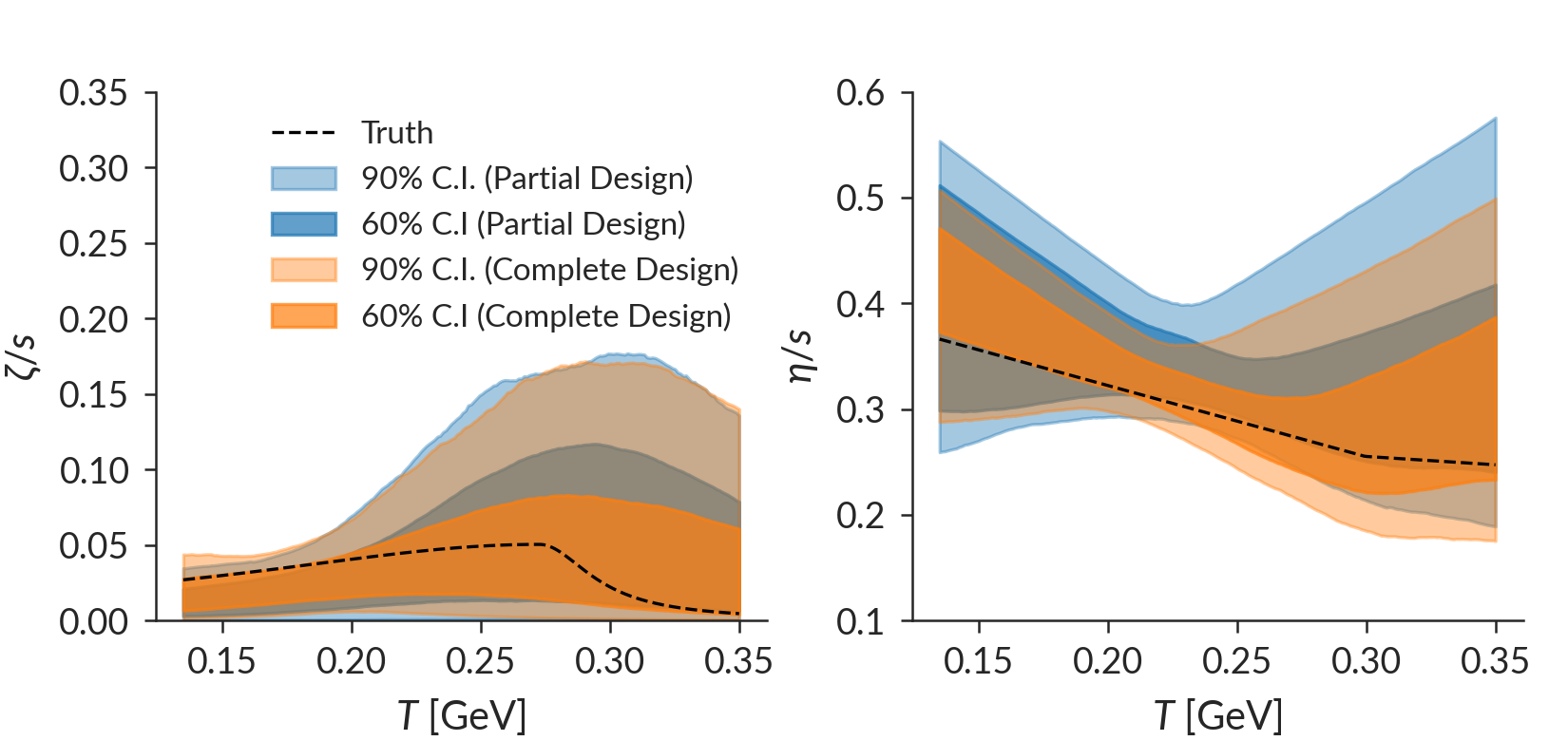}
    \caption{Viscous posterior with underlying closure test truth indicated by dashed line. The first 100 design points were used for the comparison in blue (``partial design"), while the full design was used for the comparison in orange (``full design"). }
    \label{fig:val3-viscous-posterior}
\end{figure}

A surrogate model -- in heavy ion collisions, typically Gaussian processes -- can be trained on a set of design points and be used as an inexpensive model emulator, mapping input to output at a fraction of the computational cost of the hybrid model. Importantly, those surrogate models must be validated before they can be used.
In self-consistency (or closure) testing, given model output with known input, the surrogate model can solve the inverse problem and recover the input. 
For a fixed validation point, where the model is compared to pseudodata generated with known parameters, the viscous posterior is shown in Fig.~\ref{fig:val3-viscous-posterior}. 
By increasing the number of design points in the training set, the posterior for both $\zeta/s$ and $\eta/s$ are more constrained around the underlying truth. 
\vspace{-0.1in}
This demonstrates that, with care in choosing the design points, intermediate analyses may be performed as demonstrated in our hybrid model with IP-Glasma, reducing the time required for the production of physics output. 
This can also serve to conserve computational resources by determining if a regime of diminishing returns has been reached, in turn ensuring prudent usage of limited and costly computational resources for maximal physics results. 

\section{Acknowledgements}
This work was supported in part by the Natural Sciences and Engineering Research Council of Canada, and in part by the U.S. Department of Energy Grant no. DE-FG02-05ER41367. This research was enabled in part by support provided by Compute Canada and its regional partners Calcul Quebec, Compute Ontario, SciNet, SHARCNET, and WestGrid.
\printbibliography
\end{document}